\documentclass[aip, apm,showpacs,amsmath,amssymb,amsfonts,10pt
	,onecolumn
	,superscriptaddress,floatfix,footinbib
	]{revtex4-1}

\usepackage[utf8]{inputenc}
\usepackage{amsmath}
\usepackage{amssymb}
\usepackage{graphicx}

\usepackage{upgreek}
\usepackage{SIunits}

\usepackage{units}
\usepackage{color}

\usepackage{tabularx}

\begin{document}

\title{Dielectric tensor of monoclinic Ga$_2$O$_3$ single crystals in the spectral range $0.5 - 8.5\,\mathrm{eV}$} 
\author{C. Sturm}
\affiliation{Institut f\"ur Experimentelle Physik II, Universit\"at Leipzig, Linnéstr. 5, 04103 Leipzig, Germany}

 \author{J. Furthm\"uller}
 \affiliation{Institut f\"ur Festk\"orpertheorie und -optik, Friedrich-Schiller-Universit\"at Jena, Max-Wien-Platz 1, 07743 Jena, Germany}


\author{F. Bechstedt}
 \affiliation{Institut f\"ur Festk\"orpertheorie und -optik, Friedrich-Schiller-Universit\"at Jena, Max-Wien-Platz 1, 07743 Jena, Germany}

\author{R. Schmidt-Grund}
\affiliation{Institut f\"ur Experimentelle Physik II, Universit\"at Leipzig, Linnéstr. 5, 04103 Leipzig, Germany}

\author{M. Grundmann}
\affiliation{Institut f\"ur Experimentelle Physik II, Universit\"at Leipzig, Linnéstr. 5, 04103 Leipzig, Germany}

\begin{abstract}
The dielectric tensor of $\beta$-Ga$_2$O$_3$ was determined by generalized spectroscopic ellipsometry in a wide spectral range from $0.5\,\mathrm{eV}$ to $8.5\,\mathrm{eV}$ as well as by calculation including quasiparticle bands and excitonic effects. The dielectric tensors obtained by both methods are in excellent agreement with each other and the observed transitions in the dielectric function are assigned to the corresponding valence bands. It is shown that the off-diagonal element of the dielectric tensor reaches values up to $|\varepsilon_{xz} | \approx 0.30 $ and cannot be neglected. Even in the transparent spectral range where it is quite small ($|\varepsilon_{xz} | < 0.02 $) it causes a rotation of the dielectric axes around the symmetry axis of up to $20^\circ$. 
\end{abstract}

\maketitle
\section{Introduction}

Transparent conductive oxides (TCO) are important materials for the realization of transparent electrodes \cite{Bright2007, Liu2010, Klimm2014}, e.g. for solar cells \cite{Calnan2010} and transparent transistors \cite{Frenzel2013}. Indium tin oxide (ITO) which is up to now the material of choice in industrial applications has the drawback that indium is quite expensive and therefore much effort is devoted to develop indium free TCO \cite{Liu2010}. One promising material is Ga$_2$O$_3$ with a reported band band gap energy of about $4.8\,\mathrm{eV}$ \cite{Ueda1997,Yamaga2011} so that even in the presence of crystal imperfection high transmission in the visible and even in the UV-A/B spectral range can be obtained. Furthermore, the band gap energy is larger then the absorption edge of the atmosphere so that Ga$_2$O$_3$ is also a promising candidate for solar blind photodetectors \cite{Ji2006,Guo2014}. 

At ambient conditions, Ga$_2$O$_3$ crystallizes in a monoclinic crystal structure ($\beta$-phase) and the dielectric function is a tensor rather than a scalar function. Although the optical characterization of thin films was done by several groups \cite{Kim1987,Rebien2002,Al-Kuhaili2003,Ji2006, Lv2012, Goyal2014, Mi2014}, rotation domains present in the thin films and the inferior crystal structure compared to a single crystal prohibited the determination of the tensor components and only an effective dielectric function was measured. Experiments  on single crystals are rare and mainly limited to transmission \cite{Ueda1997,Galazka2010} and absorption \cite{Matsumoto1974, Ueda1997, Yamaga2011} measurements. From such data the entire dielectric tensor cannot be determined and only the absorption edges as well as the refractive index in the transparent spectral range were deduced. This strongly limits the comparison with theoretical calculations of the band structure. Besides experimental work, the band structure as well as the tensor of the dielectric function were calculated by several authors \cite{He2006a,Yamaga2011,Varley2015}. However, for the reported dielectric function the crystal was treated as an orthorhombic system, i.e. the dielectric tensor consists only of diagonal elements and the off-diagonal element was neglected.

In this work we present the full dielectric tensor of $\beta$-Ga$_2$O$_3$ within the spectral range $0.5 - 8.5\,\mathrm{eV}$. We show that the off-diagonal element ($\varepsilon_{xz}$) has a significant impact on the orientation of the dielectric axes. This knowledge is important for polarization sensitive applications. The paper is organized as follows: first we describe the experimental setup, the investigated sample and present the theoretical framework which we used for the determination of the dielectric tensor. In the following section, the obtained dielectric tensor and the rotation of the dielectric axes are discussed. 

\section{Experiment and Theory}

\subsubsection{Sample}

\begin{figure}[b]
	\centering
	\includegraphics[width = .3\columnwidth]{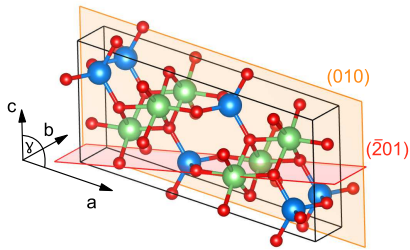}
	\caption{Schematic representation of the unit cell of $\beta$-Ga$_2$O$_3$. The tetrahedrally coordinated Ga atoms are shown in blue whereas the octahedrally coordinated ones are shown in green. The oxygen atoms are marked in red. The $(010)$ and the $(\bar{1}02)$ planes are highlighted by the orange and red rectangle, respectively. (Image created by VESTA\cite{Vesta})}
	\label{fig:schematic_ga2o3}
\end{figure}

The equilibrium crystal structure of Ga$_2$O$_3$ at ambient conditions is monoclinic, the so-called $\beta$-phase. The unit cell of this structure consists of 8 Ga atoms which are surrounded by oxygen atoms where 4 Ga atoms exhibit an octahedral, the other four a tetrahedral coordination \cite{Geller1960}. The C$_\mathrm{2h}$ symmetry axis of this structure coincides with the crystallographic $b$-axis in real space and the lattice constants are about $1.22\,\mathrm{nm}$, $0.30\,\mathrm{nm}$ and $0.58\,\mathrm{nm}$ for the $a$, $b$ and $c$-axis, respectively. The angle $\gamma$ between the $a$ and $c$-axis was determined to be $103.7^\circ$ \cite{Geller1960}. A schematic representation of the unit cell is shown in Fig.~\ref{fig:schematic_ga2o3}. 

For the investigations presented here, we have used two bulk single crystals (size of $10 \times 15\,\mathrm{mm}^2$) with different surface orientation, namely $(\bar{2}01)$ and $(010)$. They were fabricated by the Tamura Corporation by means of floating zone technique. The surface planes of the two crystals are marked in Fig.~\ref{fig:schematic_ga2o3}. The FWHM of the $(102)$ X-ray diffraction peak was specified to be $89\arcsecond$ ($48\arcsecond$) whereas the corresponding peak of the in-plane direction of the $(102)$ plane is about $46\arcsecond$ ($25\arcsecond$) for the $(010)$ ($(\bar{2}01)$) oriented sample. The as-received surface was investigated by atomic force microscopy (AFM) which revealed a smooth surface without atomic steps. The roughness was determined to be $R_\text{s}\approx 0.2\,\mathrm{nm}$ ($0.1\,\mathrm{nm}$). This smooth surface and the absence of the atomic steps indicates an amorphous surface layer with possibly different optical properties than the bulk crystal.

\subsubsection{Dielectric tensor}

The monoclinic structure of Ga$_2$O$_3$ causes an optically biaxial anisotropic behaviour with a refractive index and absorption edge depending on the polarization. In the case of Ga$_2$O$_3$ this dependence of the absorption edge on the polarization was explained by the symmetry of the valence bands and the conduction band by means of band structure calculation \cite{Yamaguchi2004,He2006, Varley2015}. These calculations also yield, that most of the charges corresponding to the conduction band minimum are located in interstitial regions whereas in the case of the valence bands the charges are located at the p-orbitals of the O atoms. The width of the valence bands was estimated to be about $7\,\mathrm{eV}$. Surprisingly, it was found that the conduction band minimum is nearly isotropic. Thus the anisotropy of the reduced mass is determined by the anisotropic behaviour of the valence bands \cite{Yamaga2011}.

The optical response of a material is described by its dielectric function (DF). For a  monoclinic system this DF is a tensor given by \cite{BornWolf}:
\begin{equation}
  \varepsilon = 
 \begin{pmatrix}
	\varepsilon_\text{xx} & 0 & \varepsilon_\text{xz}\\
	0 & \varepsilon_\text{yy} & 0 \\
	\varepsilon_\text{xz} & 0 & \varepsilon_\text{zz}
 \end{pmatrix}\,.
 \label{eq:df_tensor_monoclinic} 
\end{equation}
The $y$-direction coincides with the symmetry axis. Since this is a symmetric tensor, it can be diagonalized for each photon energy separately and the corresponding diagonal elements are the semi principal axes of the ellipsoid of wave normals \cite{BornWolf}. These axes of the ellipsoid are also often called dielectric axes. The dispersion of each element of the dielectric tensor is in general independent from the others so that the orientation of the dielectric axes with respect to the sample coordinate system depends on the photon energy\footnote{For a material with an orthorhombic crystal structure (or higher symmetry) the diagonalization of the tensor can be performed for all photon energies simultaneously and therefore the dielectric axes are fixed.}. This behaviour is called colour dispersion. For the tensor given by Eq.~\eqref{eq:df_tensor_monoclinic} one dielectric axis is fixed along the $y$-direction whereas the other two rotate as a function of the photon energy in the $x-z$-plane around the $y$-axis. 

\subsubsection{Spectroscopic Ellipsometry}

Spectroscopic ellipsometry was applied for the determination of the dielectric function of $\beta$-Ga$_2$O$_3$ in the spectral range of $0.5-8.5\,\mathrm{eV}$. This method determines the change of the polarization state of an incoming light after reflection on the sample surface \cite{AzzamBashara, Fujiwara} which is usually expressed by the ratio of the complex reflection coefficients for light polarized perpendicular (s) and parallel (p) to the plane of incidence. However, in the case of an optically anisotropic material, conversion of s- into p-polarized light takes place (and vice versa) and the  change of the polarization state is then expressed by the $4 \times 4$ Mueller matrix $\boldsymbol{M}$ by $S_\text{ref} = \boldsymbol{M} S_\text{in}$ with $S_\text{in}$ ($S_\text{ref}$) being the Stokes vector of the incident (reflected) beam \cite{AzzamBashara}. For the interpretation of the data, the Mueller matrix is often described as a matrix of four $2 \times 2$ block matrices. The diagonal block matrices are then mainly related to the different reflection coefficients for the p- and s-polarized light whereas the off-diagonal block matrix describes the mentioned conversion of the polarization from s to p (and vice versa).

For the determination of the 4 independent complex quantities of the dielectric tensor (cf. Eq.~\eqref{eq:df_tensor_monoclinic}) the measurements have to be performed at different orientations of the dielectric axes with respect to the plane of incidence \cite{Jellison2011}. This was realized by using two bulk single crystals with the surface orientation mentioned above and by rotating each crystal around its surface normal by $0^\circ$, $45^\circ$, $90^\circ$, $105^\circ$, $180^\circ$, $225^\circ$, $270^\circ$ and $305^\circ$ in order to ensure a high sensitivity to each component of the dielectric tensor. The corresponding orientation of the sample system with respect to the laboratory system is then described by using the Euler angles $(\varphi, \vartheta, \psi)$ in the $yzx$-notation \cite{Goldstein2011}. For $(\varphi, \vartheta, \psi) = (0,0,0)$, the sample as well as the laboratory system coincide with each other. Since in Ga$_2$O$_3$ the angle between the crystallographic $a$ and $c$ axis is non-orthogonal, there is  no direction which defines a cartesian coordinate system in this plane. Therefore we choose in the following the system defined by: $\hat{e}_x \parallel a$-axis, $\hat{e}_y \parallel b$-axis and $\hat{e}_z = \hat{e}_x \times \hat{e}_y$, $\hat{e}_i$ being the unit vector in the $i$ direction. As angles of incidence we have chosen $60^\circ$ and $70^\circ$.

For the determination of the dielectric function of Ga$_2$O$_3$ from the recorded spectra we used a layer stack model, consisting of the semi-infinite substrate and a surface layer. Since both single crystals were fabricated by the same technique and produced by the same company, the dielectric function was assumed to be the same and analyzed simultaneously. The surface roughness of the samples was described by an effective medium approximation (EMA) \cite{Bruggeman35}. For both substrates an effective thickness of about $2\,\mathrm{nm}$ was determined reflecting the smooth surface obtained by AFM. For the calculation of the DF of this layer, for simplicity only the diagonal elements of the dielectric tensor of Ga$_2$O$_3$ were considered and mixed with the DF of void with a fraction of $50\%:50\%$.

By applying this layer model, the DF was obtained by using a Kramers-Kronig consistent wavelength-by-wavelength analysis for each energy. By doing so, the DF exhibits a non-vanishing imaginary part in the order of $10^{-3}$ in the visible spectral range. This causes a significant absorption in this spectral range which is physically wrong. This is probably caused by the unknown optical properties of the surface layer which can lead to such an effect, similar to an isotropic material where no surface layer is taken into account, and to the limitation of the used EMA model. However, due to the small thickness of this surface layer ($d_\text{EMA} \leq 2\,\mathrm{nm}$) a determination of its DF, especially its anisotropic character, is quite challenging due to the low sensitivity to such a thin layer and the main results are unaffected by the choice of the surface layer. 
To overcome this problem, the spectral range was divided into two spectral ranges: a "transparent range`` ($0.5 - 4.5\,\mathrm{eV}$) where the absorption is set to zero and the ''absorption range`` ($3.0 - 8.5\,\mathrm{eV}$) which includes the spectral range where absorption takes place. 

In the "transparent range`` ($0.5 - 4.0\,\mathrm{eV}$) the absorption was set to zero and the dielectric function can be fully described by its real part ($\varepsilon_{1,ij}$). For materials with an orthorhombic (or higher) symmetry in this case the relation $n_{i} = \sqrt{\varepsilon_{i}}$ between the components of the refractive index ($n_i$) and the dielectric function holds where $n_i$ can be described by an Cauchy approximation. In our case, due to the presence of the off-diagonal element this relationship is not valid anymore and we used 
\begin{equation}
	\varepsilon_{1,ij} = A_{ij} + \frac{B_{ij}}{\lambda^2} + \frac{C_{ij}}{\lambda^4}\,.
\label{eq:df_approximation}
\end{equation}
Note, by using this approximation a negative dispersion can be achieved as it can be occur for the off-diagonal element of the dielectric tensor which is not possible by using an Cauchy approximation. The corresponding parameters are given in Tab.~\ref{tab:cauchy_parameter}. 

The second range includes the ''absorption range`` ($3.00 - 8.65\,\mathrm{eV}$). Here the DF was obtained by applying the wavelength-by-wavelength analysis as described above. In the spectral range where both parts overlap ($3 - 4 \,\mathrm{eV}$) a good agreement between the DF deduced by the Cauchy approximation and the wavelength-by-wavelength fit was obtained. In order to obtain a single DF for the entire spectral range, the two parts of the DF were merged where a linear weighting function was applied in the overlap range. This weighting ensures a smooth onset of the imaginary part of the DF and was done in such a way, that the transparent (absorption) spectral range contributes by $100\%$ ($0\%$) to the overall DF at an energy of $E=3\,\mathrm{eV}$ whereas at $E = 4\,\mathrm{eV}$ the contribution was set to $0\%$ ($100\%$). The overall dielectric function was than smoothed by using the Savitzky-Golay algorithm.

\begin{table}
  \centering
	\begin{tabularx}{\columnwidth}{X|X|X|X}
    \hline \hline
    & A & B & C \\
     &    &  ($10^{-2}\,\upmu\mathrm{m}^2$) & ($10^{-3}\,\upmu\mathrm{m}^4$) \\ \hline
    $\varepsilon_{xx}$ & 3.479 & 4.727 & 1.753 \\ \hline 
    $\varepsilon_{yy}$ & 3.626 & 4.739 & 1.242 \\ \hline 
    $\varepsilon_{zz}$ & 3.574 & 4.882 & 2.129 \\ \hline
    $\varepsilon_{xz}$ & 0.016 & -0.185 & -0.146 \\ \hline \hline
  \end{tabularx} 
  \caption{Parameter of the tensor elements of the dielectric function of $\beta$-Ga$_2$O$_3$ in the transparent spectral range ($E \leq 4\,\mathrm{eV}$).}
  \label{tab:cauchy_parameter}
\end{table}

We note that M.~Dressel \textit{et al.} proposed a bond polarization model \cite{Dressel2008} in order to analyse triclinic pentacene. In this model it is assumed that the dipole element and therewith the polarization is oriented along the crystallographic axes. The advantage of this model is, that the off-diagonal elements of the dielectric tensor can be expressed by the principle elements and the known angles between the crystal axes, which reduces the number of unknown independent tensor elements to three. Therefore, the major axes of the indicatrix (i.e. tensor related to the real part of $\varepsilon$) coincide with those of the conductivity tensor (i.e. related to the imaginary part of $\varepsilon$). Although this model was successfully applied to pentacene \cite{Dressel2008} and slanted columnar thin films \cite{Schmidt2013} it cannot be applied in our case and in general since the major axes of the indicatrix and the conductivity tensor do not coincide with each other (see inset Fig.~\ref{fig:df}a).

\subsubsection{Theory}

In a first step we determine the ground-state atomic structure by minimization of the energy with respect to all lattice parameters and ionic positions, employing density functional theory (DFT)\cite{kohn} and the projector-augmented wave (PAW) method\cite{paw,paw2} as implemeted in the Vienna Ab-initio Simulation Package (VASP)\cite{kresse,kresse2}. In order to obtain structural data very close to experiment we used the superior AM05 gradient-corrected exchange-correlation (XC) functional\cite{am05,am09}. A plane-wave cutoff of 820~eV and a grid of $11\times 11\times 11$ {\bf k} points has been used for the structural optimization. The high plane-wave cutoff was necessary because we employ the calculated stress tensor for optimization of the lattice parameters and this stress tensor may suffer from spurious Pulay stresses due to the incompleteness of the basis set\cite{pulay}, demanding for high plane-wave cutoffs.

The electronic excitations and optical spectra of $\beta$-Ga$_2$O$_3$ are
calculated employing state-of-the-art techniques based on many-body perturbation theory (MBPT)\cite{bechbook,onidarubioreining}. We use the GW scheme proposed by Hedin\cite{hedin} for the XC self-energy, employing an HSE\cite{hse03} hybrid functional as starting point for the electronic quasi-particle band structure to calculate single-particle excitation properties. The HSE starting point is chosen because the hybrid functional mimics already in a very crude manner parts of the quasiparticle effects. Hence the GW corrections remain much smaller than for a DFT starting point justifying much better a single-shot G$_0$W$_0$ approach. For optical properties the inclusion of excitonic effects, treated within the Bethe-Salpeter equation (BSE)\cite{bechbook,onidarubioreining} framework, is crucial in order to allow a reasonable comparison with experimental spectra. We use the standard implementation of GW and BSE of VASP which employs the full frequency-dependent dielectric function without any approximations like the plasmon pole approximation\cite{hybertsen} or model dielectric functions\cite{cappellini} for the calculation of the screened Coulomb interaction $W$ and standard static screening in the BSE case\cite{bechbook}.

For both, the GW and BSE calculations, we have to limit the {\bf k}-point sampling to a grid of $8\times 8\times 8$ {\bf k} points including the Brillouin zone center $\Gamma$ point. As a plane-wave cutoff we use a value of 410~eV. For the representation of the full microscopic inverse dielectric function entering the screened Coulomb interaction $W$ in GW and BSE we can reduce the cutoff to 150~eV. All calculated optical spectra are smoothed with a broadening of 0.2~eV. Due to high computational demands we have to use a numerically very efficient time evolution scheme\cite{smith} to solve the Bethe-Salpeter equation where the calculation of the $\omega$-dependent polarizability can be considered as an initial-value problem\cite{smith}. This avoids the diagonalization of a giant pair Hamiltonian matrix with typical ranks of 300.000 to 400.000. However, one looses information about eigenvalues and eigenvectors of the BSE pair Hamiltonian matrix. Anyway, if bound exciton peaks can be observed below the fundamental gap energy one can at least estimate the bound exciton binding energies from the peak positions relative to the fundamental single-quasiparticle gap.

\section{Results}
\subsubsection{Dielectric Function}

\begin{figure}
  \centering
\includegraphics[width =.5\columnwidth]{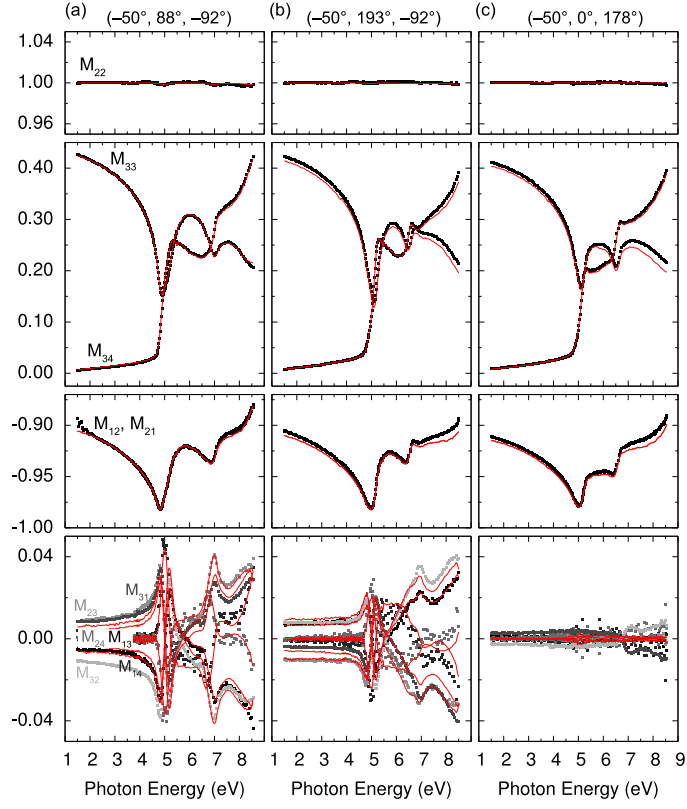}
  \caption{Experimental (symbols) and model calculated (lines) spectra of the Mueller matrix elements for an angle of incidence of $70^\circ$ for different orientations of the crystal (indicated by the Euler angles ($\varphi,\theta, \psi$) on top of each column).}
  \label{fig:mm_spectra}
\end{figure}

For selected crystal orientations with respect to the angle of incidence, the recorded as well as the calculated spectra of the Mueller matrix elements are shown in Fig.~\ref{fig:mm_spectra}. A good agreement between the experimental and the calculated spectra can be recognized. As can be seen, the off-diagonal elements of the Mueller Matrix are strongly pronounced which reflects the anisotropic character of the sample and therewith mode conversion. This effect is especially pronounced in the absorption range. Only for the case, where the symmetry axis of the crystal ($b$-axis) is perpendicular to the plane of incidence, the off-diagonal elements vanishes and no mode conversion takes place (Fig.~\ref{fig:mm_spectra}c).

\begin{figure*}
  \centering
  \includegraphics[width = 0.99\textwidth]{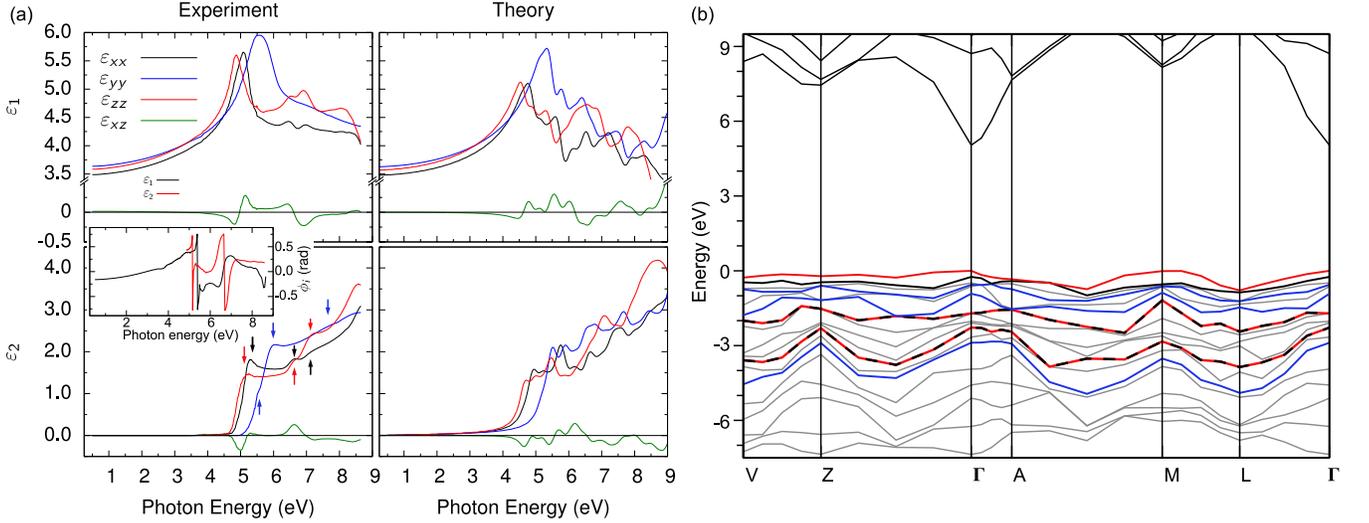}
  \caption{(a) Experimentally determined (left column) and theoretically derived (right column) tensor components of the dielectric function. inset: Rotation angle of the dielectric axes around the $y$-axis for real (black) and imaginary (red) part. (b) The calculated band structure of $\beta$-Ga$_2$O$_3$. The colour of the valence bands represents the polarization of the dipole-allowed transition into the lowest conduction band at the $\Gamma$-point (marked by arrows in (a)). The valence bands which have weakly dipole-allowed transitions are shown in light grey.}
  \label{fig:df}
\end{figure*}

The DF derived from the best match between the experimental and calculated spectra is shown in Fig.~\ref{fig:df}a. In the visible spectral range, the largest anisotropy is observed between the tensor elements for the $x$ and $y$ direction to be $\varepsilon_{yy} - \varepsilon_{xx} \approx 0.15$, whereas the difference between $\varepsilon_{yy}$ and $\varepsilon_{zz}$ is quite small ($\approx 0.05$). In the absorption range, the situation changes. In the diagonal elements, several peaks are observable caused by the transitions from valence bands into the conduction band. The first transitions occur for the dipole polarized perpendicular to the symmetry axis, namely the $z$- and $x$-direction, at around $E \approx 4.8\,\mathrm{eV}$. The first one for a polarization along the $y$-direction is shifted around $500\,\mathrm{meV}$ to higher energies. This causes a more pronounced dispersion in the transparent spectral range for the $\varepsilon_{xx}$ and $\varepsilon_{zz}$ component as for the $\varepsilon_{yy}$. At two energies, namely $E \approx 3.26\,\mathrm{eV}$ and $E \approx 4.5\,\mathrm{eV}$, the dispersion of the elements $\varepsilon_{zz}$ and $\varepsilon_{xx}$ crosses with the one of $\varepsilon_{yy}$, respectively, causing two uniaxial points.

One important property, which is as mentioned often neglected in the literature, is the monoclinic nature of $\beta$-Ga$_2$O$_3$, i.e. the presence of the off-diagonal tensor element of the DF. In the transparent spectral range ($E < 4\,\mathrm{eV}$) this element is small ($|\varepsilon_{1,\text{xz}}| < 0.02$) while in the absorption range several peaks are observable which reach values between $-0.25$ at $E \approx 4.8\,\mathrm{eV}$ and $E \approx 6.9\,\mathrm{eV}$ and $0.30$ at $E \approx 5.1\,\mathrm{eV}$. As already mentioned, the presence of the off-diagonal element causes rotation of the two dielectric axes in the $x-z$-plane around the $y$-axis by an angle $\phi$ in dependence on the energy and the optical property ($i=1,2$=real,imaginary part) given by
\begin{equation}
	\tan2\phi_i = \frac{2\varepsilon_{i,xz}}{\varepsilon_{i,xx} - \varepsilon_{i,zz}}\,.
	\label{eq:def_rotation_angle}
\end{equation}
Although the off-diagonal element in the transparent spectral range seems to be negligibly small, in conjunction with the dispersion of the diagonal elements, it causes a rotation by roughly $20^\circ$ (inset Fig.~\ref{fig:df}a). 

\subsubsection{Band structure and theoretical dielectric function}

The band structure calculated for low temperatures ($T=0$~K) in the vicinity of the upper valence bands and the lowest conduction bands is shown in Fig.~\ref{fig:df}b. By taking into account the symmetry of the bands and the corresponding transition matrix elements, the dielectric function is obtained and shown in Fig.~\ref{fig:df}a, being in excellent agreement with the experimental one. The additional peaks in the theoretical derived DF compared to the experimentally determined can be attributed to artefacts due to the limited {\bf k}-point sampling used for the calculation and the small broadening of the transitions considered in the calculations. Beside the additional peak structure, an energy shift of about $500\,\mathrm{meV}$ between the calculated and experimentally observed transition is observable. Since the experimental observed transitions are blue shifted compared to the theoretical calculated, this shift can hardly be explained by the different temperature used for the calculation and in the experiment. A reason might be that electron-phonon interaction has to be taken into account in the calculations or maybe also additional screening effects due to free carriers, although a characterization of the sample indicates free-carrier concentrations below $10^{19}$~cm$^{-3}$. Both might reduce the excitonic effects yielding a blue shift of the calculated spectra. Also strain in conjunction with a rather large deformation potential can be a potential reason. However, our calculated lattice parameters $a=1.229$~nm, $b=0.305$~nm, $c=0.581$~nm, and $\beta=103.77^\circ$ agree very well with the experimetally determined values and make this explanation  rather unlikely. In order to get a deeper understanding of this energy shift, a precise description of the DF by means of model dielectric functions including an analysis in dependence on the temperature of the corresponding oscillators is necessary which is beyond the scope of this paper.
Despite this energy shift between the theoretical and experimental DF we can, however, note that otherwise the structure of the DF and in particular also the anisotropies are reproduced perfectly. We observe the same relative shifts and ordering of the peaks and onsets for different light polarisations in Fig.~\ref{fig:df}a and also the peak heights are reproduced correctly. This means essentially, that the blue shift of the experimental spectra with respect to the theoretical ones is approximately a rigid shift.

From the calculated electronic structure and the symmetry of the bands the polarization and magnitude of the transition from the highest valence bands into the lowest conduction bands at the  $\Gamma$-point is derived. The involved valence bands which contribute mainly to the DF and the polarization of the corresponding transitions are also indicated in Fig.~\ref{fig:df}b. Except of the transitions from the first and second topmost valence bands into the conduction band which are dipole-allowed for a polarization mainly parallel to the $z$- and $x$-axis, respectively, the transitions are polarized either along the $y$-axis or within the $x-z$-plane. This is in agreement with the DF determined by means of ellipsometry where the observed transitions above the absorption edge in the tensor elements $\varepsilon_{xx}$ and $\varepsilon_{zz}$ occur at the same energy and differ from those which appear in $\varepsilon_{yy}$.

\section{Summary}

To summarize, all tensor elements of the dielectric function of $\beta$-Ga$_2$O$_3$ were determined by spectroscopic ellipsometry and calculated by modern many-body theory including quasiparticle effects in the band structure calculations as well as excitonic and local-field effects in the calculation of the dielectric tensor. They have been shown to be in excellent agreement with each other. By means of the band structure calculation, the observed transitions in the experimental dielectric function were assigned to the corresponding valence bands which explains the observed energy difference in the onset of the absorption for a dipole polarized along the symmetry axis and within the $x-z$-plane of about $500\,\mathrm{meV}$. Both, the experiment as well as the theoretical calculations, reveal a non-vanishing off-diagonal element which reaches values up to $|\varepsilon_{xz} | \approx 0.30$ in the absorption range. This element is responsible for the rotation of the dielectric axis and although it is quite small, it causes an rotation of the dielectric axes of up to $20^\circ$ in the transparent spectral range ($|\varepsilon_{xz} | < 0.02$) .

\acknowledgments
We thank Christian Kranert and Michael Bornholzer for fruitful discussions, Lennart Fricke for the AFM measurements and Carsten Bundesmann for the support of the ellipsometry measurements. This work was supported by the Deutsche Forschungsgemeinschaft within Sonderforschungsbereich 762 -  "Functionality of Oxide Interfaces". We also acknowledge financial support of the Austrian Fond zur F\"orderung der Wissenschaftlichen Forschung in the framwork of SFB25 - "Infrared Optical Nanostructures".


%

\end{document}